\documentstyle[12pt]{article}

\begin{document}
\bigskip\bigskip
\centerline{\bf HEAT KERNEL EXPANSION FOR OPERATORS IN SPACES WITH}
\centerline{\bf METRIC INCOMPATIBLE WITH CONNECTION}
\bigskip\bigskip
\centerline{\sl E.V. Gorbar, V.A. Kushnir}
\smallskip
\centerline{\sl Bogolyubov Institute for Theoretical Physics}
\centerline{\sl 252143 Kiev, Ukraine}
\bigskip\bigskip
\centerline{\bf Abstract}
\bigskip
A method
for calculation of the DWSG coefficients for operators in
spaces with metric incompatible with connection is suggested based on a
generalization of the
pseudodifferential operators technique.  By using the proposed method, the
lowest DWSG coefficients are calculated for minimal operators of the second and
fourth order and for nonminimal operators of the type $H^{\mu\nu} = - g^{\mu\nu}
g_{\alpha\beta}\nabla^{\alpha}\nabla^{\beta} + a\nabla^{\mu}\nabla^{\nu} +
X^{\mu\nu}$ in spaces with metric incompatible with connection.

\eject

\section{ Introduction}

        One of the most convenient tools for solution of various mathematical
and physical problems dealing with curved manifolds is the so called heat
kernel.  Usually it is sufficient to know only its asymptotic expansion.  The
methods to obtain this expansion are numerous and well known for
differential operators and manifolds of different kinds [1-3].  The most
popular
is that of DeWitt [1,4] which suggests a certain ansatz for heat kernel matrix
elements.  The method possesses the explicit covariance with respect to gauge
and general-coordinate transformations.  However, the DeWitt technique does not
apply to higher-order operators, nonminimal operators, operators whose leading
term is not a power of the Laplace operator, and operators defined on spaces
with metric incompatible with connection. 

	Recently,
using the Widom generalization [5] of the pseudodifferential operators
technique a new algorithm was developed [6] for computing the
DeWitt-Seeley-Gilkey (DWSG) coefficients.  The method is explicitly gauge and
geometrically covariant and allows for carrying out calculations of the
DWSG coefficients by computer [7].  As was shown in [8,9,10], the method permits
a generalization to the case of Riemann-Cartan manifolds, i.e., manifolds with
torsion, and to the case of nonminimal differential operators.  

	In this work,
by using a generalization of the method of [6], we consider the problem of
calculation of the DWSG coefficients for operators in spaces with metric
incompatible with connection. This problem arises in studying of the 
different unified scenarious dealing with dynamically generated gravity
as well as in the investigating of the effective Lagrangians of
the gauge theories and gravity (for more details see the conclusion of this
work).

 This work is built as follows. In Section 2 we calculate the lowest $E_2$ DWSG
coefficient for minimal operators of the second and fourth order and in Section 3
the lowest $E_2$ DWSG coefficient for the nonminimal operator $H^{\mu\nu} =
- g^{\mu\nu}g_{\alpha\beta}\nabla^{\alpha}\nabla^{\beta} +
a\nabla^{\mu}\nabla^{\nu} + X^{\mu\nu}$.

\section{ The DWSG coefficients for minimal operators of
the second and fourth order}

In this section we generalize the method of [6]
to the problem of calculation of the DWSG coefficients for minimal
operators of the second and fourth order in spaces with metric incompatible
with connection
\begin{equation}
A_1 = -g^{\mu\nu}\nabla_{\mu}\nabla_{\nu} + b^{\mu}\nabla_{\mu} + X,
\end{equation}
\begin{equation}
A_2 = g^{\mu\nu}g^{\alpha\beta}\nabla_{\mu}\nabla_{\nu}\nabla_{\alpha}
\nabla_{\beta} + b^{\mu\nu\alpha}\nabla_{\mu}\nabla_{\nu}\nabla_{\alpha} +
C^{\mu\nu}\nabla_{\mu}\nabla_{\nu} + d^{\mu}\nabla_{mu} + X,
\end{equation}
where X is an arbitrary matrix with respect to bundle space indices.

	For the sake of completeness, we would like to recall here the
most important steps of calculation of the DWSG coefficients in the method
proposed in [6].  In fact, the generalization of this method to the case when
operators are defined on spaces with metric incompatible with connection is
rather straightforward and when describing this method we simply indicate what
modifications are needed in the case of operators acting on spaces with
metric incompatible with connection. 

	It is well known [2,3,15] that for a
positive elliptic differential operator $A$ of order $2r$ on the n-dimensional
manifold M, the diagonal matrix elements of the heat kernel admit the
following asymptotic expansion at $t\rightarrow 0_+$:
\begin{equation}
<x|e^{-tA}|x>\simeq \sum_{m}E_m(x|A)t^{(m-n)/2r},
\end{equation}
where the summation is carried out over all non-negative integers m.
       The DWSG coefficients $E_m(x|A)$ reflect the structure of both the
operator A and the manifold M. It is the well-established fact that they are
local covariant quantities built from the coefficient functions of the
operator, curvature, torsion, and their covariant derivatives. In our case
this list includes also covariant derivatives of the metric tensor. For the
sake of simplicity, in what follows we restrict without loss of generality our 
consideration to the case $n=4$. 

	To calculate the DWSG coefficients,
we use the spectral representation of the heat kernel ${\rm exp}(-tA)$:
\begin{equation}
e^{-tA}=\int \limits_{C} \frac{id\lambda}{2\pi} e^{-t\lambda}(A-\lambda)^ {-1},
\end{equation}
where the contour $C$ goes counterclockwise around the spectrum
of the operator $A$.  This reduces our calculations to those for the resolvent
$(A - \lambda)^{-1}$.  The matrix elements of last one satisfy the following
equation:
\begin{equation}
\left(A(x,\nabla_{\mu})-\lambda\right)G(x,x^{\prime},k;
\lambda)= \frac{1}{\sqrt{g}}\delta (x - x^{\prime}).
\end{equation}
To solve it, we employ the pseudodifferential operators technique, with the
resolvent represented as
\begin{equation}
G(x,x^{\prime};\lambda)=\int \frac{d^4k}{\left(2\pi\right)^4\sqrt
{g(x^{\prime})}}\, e^{il(x, x^{\prime},k)} \sigma(x, x^{\prime},k;\lambda),
\end{equation}
and
\begin{equation}
\frac{\delta(x - x^{\prime})}{\sqrt{g}}=\int \frac{d^4k}{\left(2\pi\right)^4
\sqrt{g(x^{\prime})}}\, e^{il(x, x^{\prime},k)} I(x, x^{\prime},k;\lambda).
\end{equation}
Here $l(x, x^{\prime}, k)$ is the phase function, $I(x, x^{\prime})$ - the
so called function of parallel transport, and $\sigma(x,x^{\prime}, k;
\lambda)$ -
the amplitude [14, 15].  In the flat space $l(x, x^{\prime})$ is nothing else
as $l(x, x^{\prime}, k) = k_{\mu}(x - x^{\prime})^{\mu}$ and in the case of more
general manifolds, it is a real function, biscalar with respect to
general-coordinate
transformations.   The linearity condition in x is generalized to the
requirement for the m{\it th} symmetrized covariant derivative of l to vanish as
$x \rightarrow x^{\prime}$:
\begin{eqnarray*}
\left\{\nabla_{\mu_1}\nabla_{\mu_2}\ldots\nabla_{\mu_m}\right\}l|_
{x=x^{\prime}}=
\left[\left\{\nabla_{\mu_1}\nabla_{\mu_2}\ldots\nabla_{\mu_m}\right\}l\right] =
\end{eqnarray*}
\begin{equation}
k_{\mu_1}\,\,\, {\rm for}\ m=1\,\,\, {\rm and}\,\,\, 0\,\,\, {\rm for}\,\,\,
m\ne 1.
\end{equation}
In Eq. (7) the curly brackets denote symmetrization in all indices and the
square brackets mean that the coincidence limit $x \rightarrow x^{\prime}$ is
taken.
  The local properties of the function l are sufficient when obtaining the
diagonal heat
kernel expansion.
The biscalar function $I(x, x^{\prime})$ is defined analogously:
\begin{eqnarray*}
[{\rm I}] = 1,
\end{eqnarray*}
\begin{equation}
\left[\left\{\nabla_{\mu_1}\nabla_{\mu_2}\dots\nabla_{\mu_m}\right\}{\rm I}
\right] = 0
\,\,\,m \ge 1,
\end{equation}
the unity in Eq. (8) is generally a matrix unity.

        We first consider the case of the second order operator $A_1$ in
Eq. (1).  It follows
from Eq. (4) that the amplitude $\sigma$ satisfies the equation
\begin{eqnarray*}
(-g^{\mu\nu}(\nabla_{\mu}+i\nabla_{\mu}l)(\nabla_{\nu} + i\nabla_{\nu}l) +
b^{\mu}(\nabla_{\mu} + i\nabla_{\mu}l)
\end{eqnarray*}
\begin{equation}
+ X -\lambda)
\sigma(x,x^{\prime},k;\lambda)= {\rm I}(x,x^{\prime}).
\end{equation}
        To generate expansion (2), we introduce an auxiliary parameter
$\epsilon$ into Eq. (9) according to the rule
$l \rightarrow l/\epsilon$, $\lambda \rightarrow \lambda / \epsilon^{2},$
and expand the amplitude into a formal series in powers of $\epsilon$
\begin{equation}
\sigma(x,x^{\prime},k;\lambda)=\sum_{m=0}^{\infty}\epsilon^{2+m}
\sigma_m(x,x^{\prime},k;\lambda)
\end{equation}
(the parameter $\epsilon$ is then set equal to one).  

	In such a way Eq. (10) gives us
the recursion equations to determine the coefficients $\sigma_m$ from, and,
eventually,
this procedure leads to expansion (2) where the DWSG coefficients $E_m(x|A)$ are
expressed through the integrals of $[\sigma_m]$ in the form [6]:
\begin{equation}
E_m(x|A)= \int \frac{d^nk}{(2\pi)^n\sqrt{g}} \int \limits_C \frac{id\lambda}
{2\pi} e^{-\lambda} [\sigma_m](x,x,k;\lambda).
\end{equation}

        In the case of the second order operator $A_1$ the recursion equations
for the lowest $\sigma_0, \sigma_1$, and $\sigma_2$ coefficients take the form
\begin{eqnarray*}
(g^{\mu\nu}\nabla_{\mu}l\nabla_{\nu}l - \lambda)\sigma_0 = I,
\end{eqnarray*}
\begin{eqnarray*}
(g^{\mu\nu}\nabla_{\mu}l\nabla_{\nu}l - \lambda)\sigma_1 +
(-2ig^{\mu}{\nu}\nabla_{\mu}l\nabla_{\nu} - ig^{\mu\nu}\nabla_{\mu\nu}l +
ib^{\mu}\nabla_{\nu}l)\sigma_0 = 0,
\end{eqnarray*}
\begin{eqnarray*}
(g^{\mu\nu}\nabla_{\mu}l\nabla_{\nu}l - \lambda)\sigma_2 +
(-2ig^{\mu}{\nu}\nabla_{\mu}l\nabla_{\nu} - ig^{\mu\nu}\nabla_{\mu\nu}l +
ib^{\mu}\nabla_{\nu}l)\sigma_1 +
\end{eqnarray*}
\begin{equation}
(-g^{\mu\nu}\nabla_{\mu}\nabla_{\nu} + b^{\mu}\nabla_{\mu} + X)\sigma_0 = 0.
\end{equation}
Hence, we find
\begin{eqnarray*}
\sigma_0 = \frac{I}{\nabla^{\mu}l\nabla_{\mu}l - \lambda},
\end{eqnarray*}
\begin{eqnarray*}
\sigma_1 = \sigma_0(2ig^{\mu}{\nu}\nabla_{\mu}l\nabla_{\nu} + ig^{\mu\nu}
\nabla_{\mu\nu}l - ib^{\mu}\nabla_{\nu}l)\sigma_0,
\end{eqnarray*}
\begin{eqnarray*}
\sigma_2 = \sigma_0(2ig^{\mu}{\nu}\nabla_{\mu}l\nabla_{\nu} + ig^{\mu\nu}
\nabla_{\mu\nu}l - ib^{\mu}\nabla_{\nu}l)\sigma_1 +
\end{eqnarray*}
\begin{equation}
\sigma_0(g^{\mu\nu}
\nabla_{\mu}\nabla_{\nu} - b^{\mu}\nabla_{\mu} - X)\sigma_0.
\end{equation}
        Perfoming direct algebraic calculations, 
from (14) and (7) we get the lowest $E_{2}$ coefficient for the operator $A_{1}$
\begin{eqnarray*}
E_2 = \frac{1}{(4\pi)^2}(\frac{R}{6} - X + {\nabla_{\mu}}{b^{\mu}} -
\frac{b^2}{4} -
\frac{g_{\alpha\beta}\nabla^{\mu}\nabla_{\mu}g^{\alpha\beta}}{12}
\frac{g^{\nu\beta}\nabla_{\mu}\nabla_{\nu}g^{\mu\beta}}{3} +
\end{eqnarray*}
\begin{eqnarray*}
\frac{g_{\mu\nu}g_{\alpha\beta}\nabla^{\kappa}g^{\mu\nu}\nabla_{\kappa}
g^{\alpha\beta}}{48} +
\frac{g_{\mu\alpha}g_{\nu\beta}\nabla^{\kappa}g^{\mu\nu}\nabla_{\kappa}
g^{\alpha\beta}}{24} -
\frac{g_{\alpha\beta}\nabla_{\mu}g^{\mu\nu}\nabla_{\nu}
g^{\alpha\beta}}{12} +
\end{eqnarray*}
\begin{equation}
\frac{g_{\mu\beta}\nabla_{\alpha}g^{\mu\nu}\nabla_{\nu}
g^{\alpha\beta}}{12} -
\frac{g_{\nu\beta}\nabla_{\mu}g^{\mu\nu}\nabla_{\alpha}
g^{\alpha\beta}}{4}),
\end{equation}

where we use (see [6])
\begin{eqnarray*}
\int\frac{d^nk}{(2\pi)^n\sqrt{g}}k_{\mu_1}k_{\mu_2}\ldots k_{\mu_{2s}}f(k^2) =
\end{eqnarray*}
\begin{equation}
g_{(\mu_1\mu_2\ldots\mu_{2s})} \frac{1}{(4\pi)^{n/2}2^s\Gamma(n/2
+ s)} \int_{0}^{\infty}dk^2(k^2)^{(n-2)/2 + s} f(k^2),
\end{equation}
and $g_{(\mu_1\mu_2\ldots\mu_{2s})}$ is the symmetrized sum of
metric tensor products.  Evidently, for the space with metric compatible
with connection, i.e., if $\nabla^{\mu}g_{\alpha\beta} = 0$ we restore the
well-known result $E_2 = \frac{1}{(4\pi)^2}(\frac{R}{6} - X)$.

        For the operators of the fourth order $A_2$, the equation for the
amplitude takes 
the similar but some more cumbersome form
\begin{eqnarray*}
(g^{\mu\nu}g^{\alpha\beta}(\nabla_{\mu} + i\nabla_{\mu}l)
(\nabla_{\nu} + i\nabla_{\nu}l)
(\nabla_{\alpha} + i\nabla_{\alpha}l)(\nabla_{\beta} +
\end{eqnarray*}
\begin{eqnarray*}
i\nabla_{\beta}l) +
b^{\mu\nu\alpha}(\nabla_{\mu} +
i\nabla_{\mu}l)(\nabla_{\nu} + i\nabla_{\nu}l)
(\nabla_{\alpha} + i\nabla_{\alpha}l) +
\end{eqnarray*}
\begin{equation}
C^{\mu\nu}(\nabla_{\mu} +
i\nabla_{\mu}l)(\nabla_{\nu} + i\nabla_{\nu}l) + d^{\mu}\nabla_{\mu} +
X - \lambda)\sigma = I.
\end{equation}
        Once again, to generate expansion (2) we introduce an auxiliary
parameter
$\epsilon$ into Eq. (19) according to the rule
$l \rightarrow l/\epsilon$, $\lambda \rightarrow \lambda / \epsilon^{4},$
and expand the amplitude into a formal series in powers of $\epsilon$
\begin{equation}
\sigma(x,x^{\prime},k;\lambda)=\sum_{m=0}^{\infty}\epsilon^{4+m}
\sigma_m(x,x^{\prime},k;\lambda).
\end{equation}
Similarly to the case of the operator $A_1$, we find the lowest $E_2$ DWSG
coefficient for the operator $A_2$
\begin{eqnarray*}
E_2 = \frac{\sqrt{\pi}}{(4\pi)^2}\frac{1}{2}(\frac{R}{6} + \frac{C^{\mu}
_{\mu}}{8} - \frac{9b_{\mu}^{}{\mu\nu}b_{\nu\alpha}^{\alpha} + 6b^{\mu\nu\alpha}
b_{\mu\nu\alpha}}{1024} -
\end{eqnarray*}
\begin{eqnarray*}
\frac{3g_{\alpha\beta}\nabla_{\mu}b^{\mu\nu\alpha}}{32} +
\frac{15b_{\nu\alpha}^{\alpha}\nabla_{\mu}g^{\mu\nu}}{128} +
3g_{\alpha\beta}b^{\mu\nu}_{\nu}\nabla_{\mu}g^{\alpha\beta} +
\end{eqnarray*}
\begin{eqnarray*}
\frac{6b^{\mu}_{\nu\alpha}\nabla_{\mu}g^{\nu\alpha}}{256} +
5g_{\rho\kappa}\nabla^{\mu}g^{\rho\kappa}g_{\alpha\beta}\nabla_{\mu}
g^{\alpha\beta} +
\frac{10g_{\alpha\rho}g_{\beta\kappa}\nabla_{\mu}g^{\alpha\beta}\nabla^{\mu}
g^{\rho\kappa}}{768} -
\end{eqnarray*}
\begin{eqnarray*}
\frac{41g_{\alpha\beta}\nabla_{\mu}g^{\mu\nu}\nabla_{\nu}g^{\alpha\beta}}{192}
- \frac{g_{\beta\kappa}\nabla_{\mu}g^{\alpha\beta}\nabla_{\alpha}g^{\mu\kappa}}
{192} - \frac{17g_{\nu\beta}\nabla_{\mu}g^{\mu\nu}\nabla_{\alpha}g^{\alpha\beta}
}{64}
\end{eqnarray*}
\begin{equation}
- \frac{g^{\mu\nu}g_{\alpha\beta}\nabla_{\mu}\nabla_{\nu}g^{\alpha\beta}}
{24} + \frac{7\nabla_{\mu}\nabla_{\nu}g^{\mu\nu}}{24}).
\end{equation}

\section{The DWSG coefficients for the nonminimal operator}

In this section we calculate the lowest DWSG coefficient for the nonminimal
operator $H^{\mu\nu} = - g^{\mu\nu}g_{\alpha\beta}\nabla^{\alpha}\nabla^{\beta} +
a\nabla^{\mu}\nabla^{\nu} + X^{\mu\nu}$ in space with metric incompatible with
connection.  Operators of this type arise naturally under the quantization of
gauge and gravitational fields in arbitrary gauges [11].  In [9] the lowest
DWSG coefficients were calculated by using a generalization of the
pseudodifferential operators technique.  Here, we calculate the lowest $E_2$
coefficient for the operator $H^{\mu\nu}$ in space with metric incompatible
with connection.

The most essential point as compared to the case of minimal operators
consists in alteration of recursion relations for the amplitude of resolvent
$(H - \lambda)^{-1}$ (see [9]).  The equation for the amplitude
$\sigma_{\rho\nu}(x,x^{\prime},k;\lambda)$ has the form
\begin{eqnarray*}
(g^{\mu\rho}(\nabla^{\kappa}l\nabla_{\kappa}l - i\nabla^{\kappa}\nabla_
{\kappa}l - 2i\nabla^{\kappa}l\nabla_{\kappa} - \nabla^{\mu}\nabla_{\mu} -
\lambda) + a(i\nabla^{\mu}\nabla^{\rho}l
\end{eqnarray*}
\begin{equation}
- \nabla^{\mu}l\nabla^{\rho}l +
i\nabla^{\mu}l\nabla^{\rho} + i\nabla^{\rho}l\nabla^{\mu} + \nabla^{\mu}
\nabla^{\rho}) +X^{\mu\rho})\sigma_{\rho\nu} = I^{\mu}_{\nu}(x,x^{\prime}).
\end{equation}
Setting $l \rightarrow l/\epsilon$, $\lambda \rightarrow \lambda /
\epsilon^{2},$ and $\sigma(x,x^{\prime},k;\lambda)=\sum_{m=0}^{\infty}
\epsilon^{2+m}\sigma_m(x,x^{\prime},k;\lambda)$, we get from Eq. (25) the
recursion relations for the coefficients $\sigma_{m\rho\nu}$
\begin{eqnarray*}
D^{\mu\rho}\sigma_{0\rho\nu} = I^{\mu}_{\nu}
\end{eqnarray*}
\begin{eqnarray*}
D^{\mu\rho}\sigma_{1\rho\nu} + i(-g^{\mu\nu}(\nabla_{\kappa}\nabla^{\kappa}
l + 2\nabla^{\kappa}l\nabla_{\kappa}) + a(\nabla^{\mu}\nabla^{\rho}l +
\end{eqnarray*}
\begin{eqnarray*}
\nabla^{\mu}l\nabla^{\rho} + \nabla^{\rho}l\nabla^{\mu}))\sigma_{0\rho\nu}
= 0
\end{eqnarray*}
\begin{eqnarray*}
D^{\mu\rho}\sigma_{m\rho\nu} + i(-g^{\mu\nu}(\nabla_{\kappa}\nabla^{\kappa}
l + 2\nabla^{\kappa}l\nabla_{\kappa}) + a(\nabla^{\mu}\nabla^{\rho}l +
\nabla^{\mu}l\nabla^{\rho} +
\end{eqnarray*}
\begin{equation}
\nabla^{\rho}l\nabla^{\mu}))
\sigma_{m-1\rho\nu} +
(- g^{\mu\rho}g_{\alpha\beta}\nabla^{\alpha}\nabla^{\beta}
+ a\nabla^{\mu}\nabla^{\rho} + X^{\mu\rho})\sigma_{m-2\rho\nu} = 0, m \ge 2.
\end{equation}

The main difference from the case of minimal operators is that for obtaining
$\sigma_{m\rho\nu}$ we must now invert the matrix $D^{\mu\rho}$ and
differentiate it.  Of course, this increases the algerbaic labour bot does not
cause essential diffficulties.  Solving Eq. (27) and using the
formula for the integration in k and $\lambda$ (see [9])
\begin{eqnarray*}
\int\frac{d^nk}{(2\pi)^n\sqrt{g}}(k^2)^pk_{\mu_1}k_{\mu_2}\ldots k_{\mu_{2s}}
\int_C\frac{id\lambda}{2\pi} \frac{e^{-\lambda}}{(k^2 - \lambda)^q((1-a)k^2 -
\lambda)^m} =
\end{eqnarray*}
\begin{equation}
g_{(\mu_1\mu_2\dots\mu_{2s})} \frac{\Gamma(n/2 + s + p)
F(m,p + s + n/2,q + m;a)}{(4\pi)^{n/2}2^s\Gamma(n/2 + s)\Gamma(l + m)},
\end{equation}
where $F(a,b,c;z)$ is the Gauss hypergeometric function, we obtain the
lowest $E_{2}$ DWSG coefficient
\begin{eqnarray*}
E_{2\mu\nu} = \frac{1}{(4\pi)^2}(g_{\mu\nu}R(\frac{1}{6} +
\frac{a^3 - 3a^2 + 2a}{24(1 - a)^3}) + R_{\mu\nu}\frac{9a^3 - 2a^2 + 12a}
{36(1 - a)^3} +
\end{eqnarray*}
\begin{eqnarray*}
W_{\mu\nu}\frac{a(2a - 1)}{2(1 - a)^2} -
(X_{\mu\nu} - X_{\nu\mu})(\frac{1}{2} + \frac{a}{4(1 - a)}) -
g_{\mu\nu}X^{\alpha}_{\alpha}\frac{a^2}{24(1 - a)^2} +
\end{eqnarray*}
\begin{eqnarray*}
\frac{g_{\mu\nu}g_{}{\kappa\rho}T^{\kappa\rho}_{\alpha\beta}}{6(1 - a)} +
\frac{g_{\mu\nu}T^{\kappa\rho}_{\kappa\rho}(a^2 + 4)}{3(1 - a)} -
\frac{g_{\mu\rho}T^{\kappa\rho}_{\alpha\beta}(5a^2 + 2a + 1)}{6(1 - a)} -
\end{eqnarray*}
\begin{eqnarray*}
\frac{g_{\mu\rho}T^{\kappa\rho}_{\kappa\beta}(2a^2 + a)}{1 - a} +
\frac{g_{\kappa\nu}T^{\kappa\rho}_{\mu\rho}(a^2 + 4a - 1)}{3(1 - a)} +
\frac{g_{\kappa\rho}T^{\kappa\rho}_{\mu\nu}(a^2 + 3a - 3)}{3(1 - a)} +
\end{eqnarray*}
\begin{eqnarray*}
\frac{g_{\mu\kappa}g_{\rho\nu}g^{\alpha\beta}T^{\kappa\rho}_{\alpha\beta}
(-5a^3 + 8a^2 - a + 2)a}{6(1 - a)^2} +
\end{eqnarray*}
\begin{eqnarray*}
\frac{g_{\mu\kappa}T^{\kappa\rho}_{\rho\nu}a(-2a^3 + 2a^2 + 14a - 13)}
{6(1 - a)^2} +
\end{eqnarray*}
\begin{eqnarray*}
\frac{g_{\rho\nu}T^{\kappa\rho}_{\mu\kappa}(a^2 + 2)}{3(1 - a)} +
\frac{g_{\mu\rho}T^{\kappa\rho}_{\kappa\nu}a(2 - a)}{1 - a} +
\end{eqnarray*}
\begin{eqnarray*}
g^{\alpha\beta}Q^{\kappa}_{\kappa\mu}Q_{\nu\alpha\beta}(-\frac{a}{3} +
\frac{15a^3 + 2a(a - 2)(5a^2 + 45a + 26)}{30(1 - a)^2}) +
\end{eqnarray*}
\begin{eqnarray*}
Q^{\kappa}_{\mu\nu}Q^{\rho}_{\rho\kappa}(4  + \frac{6a}{1 - a} - \frac{a(a-2)
(a^2 + 7a - 7)}{6(1 - a)^2} +
\end{eqnarray*}
\begin{eqnarray*}
\frac{a(a - 2)(5a^2 + 45a + 26)}{15(1 -a )^2} +
\end{eqnarray*}
\begin{eqnarray*}
Q^{\kappa}_{\kappa\mu}Q^{\rho}_{\rho\nu}(-\frac{a(a - 2)}{6} + \frac{72a -
4a^3 - a(a-2)(a+14) - 16(6a^2+a+4)}{12(1-a)} +
\end{eqnarray*}
\begin{eqnarray*}
\frac{15a^3 + a(a-2)
(5a^2+45a+26)}{15(1-a)^2}) +
\end{eqnarray*}
\begin{eqnarray*}
g^{\alpha\beta}Q^{\kappa}_{\alpha\beta}Q_{\kappa\mu\nu}(\frac{8a + 3}{3(1-a)} -
\frac{a(a-2)(5a^2+45a+26)}{30(1-a)^2}) +
\end{eqnarray*}
\begin{eqnarray*}
Q^{\kappa}_{\mu\nu}Q^{\rho}_{\rho\kappa}(4  + \frac{6a}{1 - a} - \frac{a(a-2)
(a^2 + 7a - 7)}{6(1 - a)^2} +
\end{eqnarray*}
\begin{eqnarray*}
\frac{a(a - 2)(5a^2 + 45a + 26)}{15(1 -a )^2} +
\frac{a^2}{3(1 - a)} + \frac{a^3}{2(1 - a)^2} - \frac{a^2}{2}) +
\end{eqnarray*}
\begin{eqnarray*}
g^{\alpha\beta}Q^{\kappa}_{\kappa\beta}Q_{\nu\mu\alpha}(\frac{a(2a^2 + a + 32)}
{6(1 - a)} + \frac{15a^3 + 2a(a-2)(5a^2 + 45a + 26)}{30(1-a)^2}) +
\end{eqnarray*}
\begin{eqnarray*}
g^{\alpha\beta}Q^{\kappa}_{\alpha\beta}Q_{\nu\mu\kappa}(\frac{a(2a^2 - 5a + 10)}
{12(1-a)} + \frac{a(a-2)25a^2+185a+104}{120(1-a)^2}) +
\end{eqnarray*}
\begin{eqnarray*}
Q^{\kappa}_{\mu\rho}Q^{\rho}_{\nu\kappa}(2a + \frac{a(a-2)(85a^2 + 465a+ 68)}
{120(1-a)^2}) +
\end{eqnarray*}
\begin{eqnarray*}
g^{\alpha\beta}Q^{\kappa}_{\mu\alpha}Q_{\nu\beta\kappa}(- \frac{2a}{3} +
\frac{a(2a^2 - a + 32)}{6(1-a)} + \frac{a(a-2)(15a^2 + 185a + 108)}
{60(1-a)^2}) +
\end{eqnarray*}
\begin{eqnarray*}
g^{\alpha\beta}Q^{\kappa}_{\kappa\nu}Q_{\mu\alpha\beta}(2a^2 - \frac{a(a+2)}{12}
- \frac{a}{3} -
\end{eqnarray*}
\begin{eqnarray*}
\frac{a(19a^2 + 65a + 30)}{24(1 - a)} -
\frac{a(5a^3 + 35a^2 +73a + 14)}{60(1 - a)^2}) +
\end{eqnarray*}
\begin{eqnarray*}
g^{\alpha\beta}Q^{\kappa}_{\kappa\alpha}Q_{\mu\nu\beta}(\frac{a(3a^2 + 16a +
32)}{12(1 - a)} + \frac{a(75a^2 - 138a  - 64)}{30(1 - a)^2})
\end{eqnarray*}
\begin{eqnarray*}
g^{\alpha\beta}Q^{\kappa}_{\alpha\beta}Q_{\mu\nu\kappa}(\frac{a(21a^2 + 46a -
32)}{24(1-a)} + 
\end{eqnarray*}
\begin{eqnarray*}
\frac{5a^2(a^2 - 5a - 14)) + a(a - 2)(10a^2 + 120a + 74)}
{120(1-a)^2}) +
\end{eqnarray*}
\begin{eqnarray*}
g^{\alpha\beta}Q^{\kappa}_{\nu\beta}Q_{\mu\alpha\kappa}(\frac{a(a - 2)}{3} +
\frac{a(16a^2 + 11a + 26)}{12(1 - a)} +
\end{eqnarray*}
\begin{eqnarray*}
\frac{5a^2(a^2 - 3a - 8) + 2a(a-2)
(5a^2 + 60a + 37)}{60(1-a)^2}) +
\end{eqnarray*}
\begin{eqnarray*}
g^{\alpha\beta}g^{\kappa\rho}Q_{\mu\alpha\beta}Q_{\nu\kappa\rho}(\frac{a(5a^2 +
a - 1)}{12(1-a)} +\
\end{eqnarray*}
\begin{eqnarray*}
frac{a(a-2)(5a^2 + 60a + 37) - 10a^2(a + 3)}{60(1-a)^2}) +
\end{eqnarray*}
\begin{eqnarray*}
g^{\alpha\beta}g^{\kappa\rho}Q_{\mu\alpha\kappa}Q_{\nu\beta\rho}(\frac{a(5a^2 +
a - 1)}{6(1-a)} +
\end{eqnarray*}
\begin{eqnarray*}
\frac{a(a-2)(5a^2 + 60a + 37) - 10a^2(a + 3)}{60(1-a)^2}) +
\end{eqnarray*}
\begin{eqnarray*}
g_{\mu\nu}g^{\alpha\beta}g^{\sigma\tau}Q_{\kappa\alpha\beta}Q^{\kappa}_{\sigma
\tau}(-\frac{1}{6} +
\end{eqnarray*}
\begin{eqnarray*}
\frac{5a(-5a^2 - 5a + 8)}{120(1-a)^2} + \frac{a(a-2)(5a^2 +
45a + 26)}{120(1-a)^2}) +
\end{eqnarray*}
\begin{eqnarray*}
g_{\mu\nu}g^{\alpha\beta}g^{\sigma\tau}Q_{\kappa\alpha\rho}Q^{\kappa}_{\beta
\sigma}(-\frac{1}{3} +
\end{eqnarray*}
\begin{eqnarray*}
\frac{5a(-5a^2 -5a + 8) + a(a-2)(5a^2 + 45a + 26)}
{60(1-a)^2}) +
\end{eqnarray*}
\begin{equation}
g_{\mu\nu}g^{\alpha\beta}Q^{\kappa}_{\alpha\rho}Q^{\rho}_{\kappa\beta}(-2 +
\frac{a(a-2)(5a^2 + 45a + 26)}{30(1-a)^2}) +
g^{\alpha\beta}Q_{\mu\alpha\kappa}Q^{\kappa}_{\beta\nu}(\frac{-a(a -2)}{2}),
\end{equation}
where $T^{\mu\nu}_{\alpha\beta} = \nabla^{\mu}\nabla^{nu}g_{\alpha\beta}$ and
$Q_{\mu\alpha\beta} = \nabla_{mu}g_{\alpha\beta}$.

Here, we have calculated the lowest DWSG coefficients for the case of metrics
incompatible with
connection. We can encounter this problem in studying different 
unified models dealing with dynamically generated gravity. It has long been 
noted (Schr{\" o}dinger, ref. [11]) that the basic concepts of the Einstein
gravity (Riemann tensor, curvature, invariant differentiation and so on) are
not characteristic for the 
models based on the metric connection. Instead, it is more natural to consider 
a models with only fundamental affine connection. 

	Another approach leading to metric incompatible
with the connection is a hypothesis of the matter field as a source for the
metric tensor.  For instance, in [12,13] the usual Einstein action is obtained
as a result of integration over quantum fluctuations of the fundamental matter
(usually fermion) fields.
Metric tensor is then obtained as a vacuum background field. The compatibility
of this
field with the connection is viewed as dynamical equation (for example, the 
minimization of the vacuum energy leading to requirement for this background
to be covariantly constant) rather than purely geometrical relation. Note that
this approach can allow for a better ultraviolet behaviour.

        At last, the metric incompatible with connection can appear in the
effective theories when reducing the effective action to a purely quadratic
form.  Evidently, there is no reason
why the (effective) metric tensor in such a case should automatically be
compatible with a
connection.

        In work [14] it was suggested to find DWSG coefficients for operators
in spaces with torsion by using DWSG coefficients for operators without torsion.
This can be done redefinition of covariant derivative which again leads to the
problem of calculation of DWSG coefficients in spaces with metric incompatible
with connection.

\bigskip\bigskip
\section{Acknowledgments}
The authors are grateful to V.P. Gusynin for many valuable remarks
and fruitful discussions.
The work was supported in part by the grants INTAS-93-2058-EXT "East-West
network in constrained dynamical systems".

\end{document}